\documentclass{article}

\usepackage{spconf,amsmath,graphicx}
\usepackage{subfigure}
\usepackage{latexsym}
\usepackage{amssymb}
\usepackage{amsfonts}
\usepackage{epsfig}
\usepackage{cite}
\usepackage{calc}
\usepackage{color}
\usepackage{theorem}
\usepackage{url}
\usepackage{subfigure}
\usepackage{cases}
\usepackage{verbatim}
\usepackage{array}
\usepackage{dcolumn}
\usepackage{multirow}
\usepackage{tabularx}
\usepackage{setspace}
\usepackage{algorithmicx} 
\usepackage{algorithm}
\usepackage{caption}
\usepackage{algpseudocode}
\usepackage{lipsum}
\usepackage{enumitem}
\usepackage{scrextend}
\addtokomafont{labelinglabel}{\sffamily}

\def\bx{{\mathbf{x}}}

\def\b0{{\mathbf{0}}}

\newcommand{\fig}[1]{Fig.\ \ref{#1}}

\newmuskip\pFqmuskip

\def\argmin{\mathop{\rm \arg\!\min}}

\title{ADC Bit Allocation under a power constraint for\\ mmWave massive MIMO Communication Receivers}
\twoauthors
  {Jinseok Choi and Brian L. Evans 
}
	{\normalsize Wireless Networking and Communications Group\\
\normalsize The University of Texas at Austin, Austin, TX USA}
  {Alan Gatherer}
	{\normalsize Wireless Access Laboratory\\
\normalsize Huawei Technologies, Plano, TX USA}

\begin{document}
%
\maketitle

\begin{abstract}

Millimeter wave (mmWave) systems operating over a wide bandwidth and using a large number of antennas impose a heavy burden on power consumption.
In a massive multiple-input multiple-output (MIMO) uplink, analog-to-digital converters (ADCs) would be the primary consumer of power in the base station receiver. 
This paper proposes a bit allocation (BA) method for mmWave multi-user (MU) massive MIMO systems under a power constraint.
We apply ADCs to the outputs of an analog phased array for beamspace projection to exploit mmWave channel sparsity.
We relax a mean square quantization error (MSQE) minimization problem and map the closed-form solution to non-negative integer bits at each ADC.
In link-level simulations, the proposed method gives better communication performance than conventional low-resolution ADCs for the same or less power.
Our contribution is a near optimal low-complexity BA method that minimizes total MSQE under a power constraint.

\end{abstract}


\vspace{-1em}
\section{Introduction}
\label{sec:intro}
\vspace{-0.5em}

Millimeter wave communication is a promising technology for next-generation cellular systems~\cite{pi2011introduction,rappaport2013millimeter,andrews2014will,boccardi2014five}. 
To compensate the large path loss, a high beamforming gain is necessary. 
The small wavelength allows large antenna arrays with small antenna spacing and can potentially lead to orders of magnitude increase in data rates~\cite{swindlehurst2014millimeter,pi2012millimeter}.
Large antenna arrays, however, consume significant power with excessive ADC bit-rate due to a large signal bandwidth and a high number of bits/sample.


To overcome the limitations, MIMO systems with low resolution ADCs have been of interest~\cite{verenzuela2016hardware,risi2014massive, mo2015capacity,orhan2015low}. 
In~\cite{risi2014massive}, the performance of 1-bit ADC massive MIMO uplink system was evaluated for Rayleigh fading channels.  
For mmWave channels, the capacity of massive MIMO with 1-bit quantization was studied~\cite{mo2015capacity} with the channel state information known at the transmitters and the receivers (CSIR).
In~\cite{mo2014channel}, the channel estimator in mmWave MIMO systems with 1-bit ADCs was proposed by modifying the expectation maximization algorithm to solve a 1-bit compressed sensing problem.
In~\cite{choi2016near}, near maximum likelihood (nML) detector and channel estimator for uplink 1-bit ADC massive MIMO systems were introduced by relaxing the norm constraint of a ML estimator.
While frequency flat channels were assumed in~\cite{risi2014massive,mo2015capacity,mo2014channel, choi2016near}, the achievable rate of uplink 1-bit ADC massive MIMO systems in the frequency selective channel was derived in~\cite{mollen2016performance}.
A mixed ADC approach has also been studied. 
In~\cite{liang2016mixed}, the achievable rate of massive MIMO systems with 1-bit ADCs partially replaced with high resolution ADCs was analyzed by using the generalized mutual information.
Using probabilistic Bayesian inference, detectors for the massive MIMO uplink system with the mixed-ADC receiver architecture were proposed in~\cite{zhang2015mixed}.
To examine the effect of ADC resolutions, an additive quantization noise model (AQNM) was considered~\cite{orhan2015low,fan2015uplink,zhang2016spectral}. 
Under the AQNM, the effect on a point-to-point MIMO mmWave system was investigated in~\cite{orhan2015low}. 
The achievable rate of uplink massive MIMO systems with low-resolution ADCs using maximum-ratio combining was derived for Rayleigh~\cite{fan2015uplink} and Rician fading channels~\cite{zhang2016spectral}. 

In this paper, we investigate ADC bit allocation (BA) for mmWave MU-massive MIMO uplink systems.
Whereas the previous work studied low-resolution ADC massive MIMO with given bits~\cite{risi2014massive,mo2015capacity,mo2014channel,choi2016near,mollen2016performance,liang2016mixed,zhang2015mixed,orhan2015low,fan2015uplink,zhang2016spectral} or proposed a greedy BA approach under a bit constraint~\cite{gersho2012vector,choispace},
we propose a near optimal low-complexity BA algorithm for minimizing quantization error.
We adopt radio frequency (RF) preprocessing
to exploit sparsity of mmWave channels~\cite{heath2016overview} by projecting received signals onto the beamspace. 
Having the total ADC power consumption of the receiver with uniform bit ADCs as a power constraint, we formulate a relaxed MSQE minimization problem for the beamspace signals under the AQNM assumption. 
Through non-negative integer mapping, the solution of the problem is used to allocate quantization bits at each ADC.
Simulation results illustrate that our BA algorithm outperforms the low-resolution quantization system in error vector magnitude (EVM) where all ADC resolutions are the same.

\vspace{-0.7em}
\section{System and channel model}
\label{sec:sys_model}
\vspace{-0.5em}

\begin{figure}[!t]
  \centering
  \centerline{\resizebox{0.9\columnwidth}{!}{\includegraphics{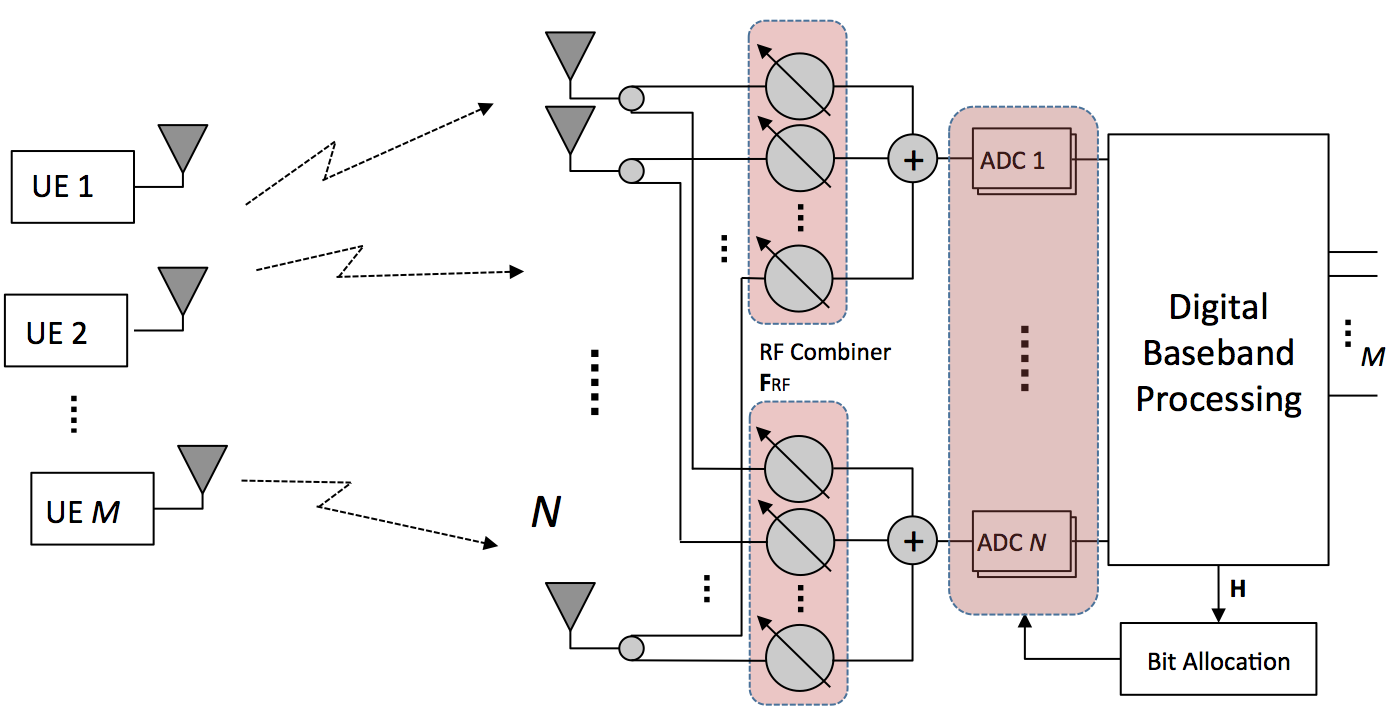}}}
\caption{The uplink multi-user massive MIMO system with $M$ user elements (UEs), $N$ antenna elements, analog beamforming and ADC bit allocation.} 
\label{fig:system}
\vspace{-0.5em}
\end{figure}

We consider MU-MIMO uplink networks in which $M$ users are equipped with a single transmit antenna. 
The base station (BS) with $N$ antennas receives the user signals, and we assume $N \gg M$.
The antenna array is a uniform linear array (ULA).
The mmWave communication uses narrowband channels with the CSIR. 
RF preprocessing is adopted at the BS so that an analog beamformer ${\bf F_{\rm RF}}  \in \mathbb{C}^{N\times N}$\footnote{We consider a square matrix as we do not aim to reduce power consumption from analog beamforming itself, but from using low-resolution ADCs.} is applied to the received signal ${\bf y} \in \mathbb{C}^N$. 
Each element of $\bf F_{\rm RF}$ is limited to have an equal norm of $1/\sqrt{N}$.
Each beamforming output is connected to an ADC pair as shown in the \fig{fig:system}.
At each ADC, a real or imaginary component of the complex signal is quantized.
The received signal after the beamforming is
\begin{align}
\label{eq:rx_signal} 
\mathbf{\tilde y}
={\bf F}^H_{\rm RF}\,\mathbf{y} 
= \mathbf{F}^H_{\rm RF}\,\mathbf{Hx} + \mathbf{F}^H_{\rm RF}\,\bf n
\end{align} 
where ${\bf H} \in \mathbb{C}^{N\times M}$ is the channel matrix, ${\bf x} \in \mathbb{C}^{M} $ is the vector of $M$ user symbols and ${\bf n} \in \mathbb{C}^{N}$ is the additive white Gaussian noise which follows complex Gaussian distribution $\mathcal{CN}(\mathbf{0}, N_0\mathbf{I}_N)$. 
We assume $\mathbb{E}[x_i] = 0$ and $\mathbb{E}[|x_i|^2] = 1$.

Since mmWave channels are expected to have limited scattering~\cite{heath2016overview}, the $i$th user channel $\mathbf{h}_i \in \mathbb{C}^{N}$ is assumed to be a sum of the contributions of $p$ scatterings.
To represent $\bf H$ in beamspace, we define the matrix $\mathbf{A}=[{\bf a}(\theta_1),\cdots,{\bf a}(\theta_N)]$. 
The vector ${\bf a}(\theta_i)$ is the array response vector where $\theta_i \in [0,2\pi]$ is the azimuth angles of arrival.
Adopting the virtual channel representation~\cite{sayeed2002deconstructing,mendez2016hybrid}, $\bf H$ can be modeled as
\begin{align} 
\nonumber
{\bf H} &= \sqrt{N/p}\,[{\bf a}(\theta_1),\cdots,{\bf a}(\theta_N)][{{\bf h}_{\rm b}}_1,\cdots,{{\bf h}_{\rm b}}_M]\\ \label{eq:channel}
&=  \sqrt{N/p}\,\bf AH_{\rm b} = {\bf AG}
\end{align}
where ${{\bf h}_{\rm b}}_i\in \mathbb{C}^{N}$ is the beamspace channel of the $i$th user
and $ {\bf G} =  \sqrt{N/p}\,\bf H_{\rm b} $.
Note that ${{\bf h}_{\rm b}}_i$ has $p$ channel gains corresponding to $p$ scatterings and the other $(N-p)$ channel gains are assumed to be much smaller than the $p$ channel gains.
Under the ULA assumption with uniformly spaced spatial angles $\vartheta_i = i/N$, $\bf A$ becomes the Fourier transform matrix~\cite{heath2016overview}; ${\bf a}(\theta_i) = \frac{1}{\sqrt{N}}\left[1,e^{-j2\pi\vartheta_i},\cdots,,e^{-j2\pi\vartheta_i(N-1)}\right]^\intercal$ with $\theta_i = \arcsin(\frac{\lambda\vartheta_i}{d})$ where $\lambda$ is the signal wavelength and $d$ is the distance between antenna elements.

The AQNM~\cite{orhan2015low} with different quantization bits for each symbol is considered so that the quantized signal of $\mathbf{\tilde y}$ is
\begin{align} 
 \label{eq:AQNM} 
\mathbf{\tilde y}_{\rm q} &= Q(\mathbf{\tilde y}) = \mathbf{W}_{\alpha} \,\mathbf{\tilde y}+ \mathbf{n}_{\rm q}
\end{align} 
$Q(\cdot)$ is a quantizer function and $\mathbf{W}_\alpha =  \text{diag}(\alpha_1,\cdots, \alpha_N)$ is a diagonal matrix with  $\alpha_i = 1- \beta_i$. 
Here, $\beta_i$ is the ratio of the MSQE and the power of the symbol, i.e., $\beta_i = \frac{\mathbb{E}[|\tilde {y}_i - \tilde{y}_{{\rm q}i}|^2]}{\mathbb{E}[|\tilde{y}_i|^2]} $ where $ \tilde{y}_{{\rm q}i}$ is the quantized output for $\tilde{y}_i$.
We consider $\mathbf{n}_{\rm q}$ as the additive quantization noise uncorrelated with $\bf \tilde y$, and $\mathbf{n}_{\rm q}$ follows the complex Gaussian distribution~\cite{fan2015uplink,zhang2016spectral}.
For a fixed channel realization $\bf H$, the covariance matrix of $\bf n_{\rm q}$ is
\begin{align} 
\label{eq:cov}
\mathbf{R}_{\mathbf{n}_{\rm q}}
= {\bf W}_\alpha {\bf W}_\beta \,{\rm diag}({\bf F}^H_{\rm RF}{\bf H}({\bf F}^H_{\rm RF}{\bf H})^H + N_0{\bf I}_{N})
\end{align}
where ${\bf W_\beta}$ is a diagonal matrix;  $\mathbf{W}_\beta =  \text{diag}(\beta_1,\cdots, \beta_N)$, and ${\rm diag}({\bf F}^H_{\rm RF}{\bf H}({\bf F}^H_{\rm RF}{\bf H})^H + N_0{\bf I}_{N})$ is also a diagonal matrix with the diagonal terms of (${\bf F}^H_{\rm RF}{\bf H}({\bf F}^H_{\rm RF}{\bf H})^H + N_0{\bf I}_{N}$).
$\bf R_{n_{\rm q}}$ is a non-Hermitian matrix as $\bf n_{\rm q}$ is uncorrelated with $\bf \tilde y$.

\vspace{-0.5em}
\section{Variable Bit Allocation in ADCs}
\label{sec:main}

In the beamspace, the channel of each user ${\bf h_{\rm b}}_i$ has $p$ major channel gains and $(N-p)$ much smaller gains, so that the channel can be regarded to be nearly sparse. 
We leverage such channel sparsity to improve the communication performance under the power constraint, and adopt the MSQE, $\mathbb{E}\left[|\tilde y_i - \tilde {y_{\rm q}}_i|^2\right]$, as a distortion measure.
Thereby, we choose $\bf F_{\rm RF} = A$, which gives the spatial Fourier transform of the received signal by projecting the signal onto the beamspace. 
For the beamspace projected sparse signals, 
we formulate the MSQE minimization problem to efficiently allocate quantization bits under the power constraint.

\vspace{-0.5em}
\subsection{RF Preprocessing: Beamspace Projection}
\label{subsec:A} 

Leveraging RF preprocessing, we project the received signal onto the beamspace with $\bf F_{\rm RF} = A$. 
The received signal after the analog beamforming~\eqref{eq:rx_signal} becomes
\begin{align} 
\label{eq:fft_signal}
{\bf \tilde{y}}  = {\bf A}^H{\bf Hx} + {\bf A}^H \bf n = \bf  Gx + \tilde n
\end{align}
where $ {\bf \tilde n =  A}^H\bf n$, and $\bf \tilde n$ follows $\mathcal{CN}(\mathbf{0}, N_0\mathbf{I}_N)$ as $\bf A$ is unitary. 
Since $\tilde{y}_i = [\mathbf{G}]_{i,:}\,\mathbf{x} + \tilde{n}_i = \sum_{u=1}^{M} g_{i,u}x_u + \tilde{n}_i$, the signal $\tilde y_i$ becomes a complex Gaussian random variable by the central limit theorem (CLT) as $M\to \infty$.
Based on the CLT, we approximate $\tilde y_i$ as a complex Gaussian random variable with the variance of $\sigma^2_{i}= [\mathbf{G}]_{i,:} [\mathbf{G}]^H_{i,:} + N_0$.

Using~\eqref{eq:fft_signal}, we can rewrite the quantized signal $\bf \tilde y_{\rm q}$ in~\eqref{eq:AQNM} as
\begin{align} 
\label{eq:AQNM2}
\mathbf{\tilde y}_{\rm q} = \bf{W_\alpha Gx + W_\alpha \tilde n + n_{\rm q}}. 
\end{align} 
Consequently, the covariance matrix of $\bf n_{\rm q}$ in~\eqref{eq:cov} becomes $\mathbf{R}_{\mathbf{n}_{\rm q}}= {\bf W}_\alpha {\bf W}_\beta \,{\rm diag}({\bf G G}^H + N_0{\mathbf{I}_{N}}).$
Assuming the non-linear scalar minimum mean square error (MMSE) quantizer, $\beta_i$ becomes $\beta_i = \frac{\pi\sqrt{3}}{2} 2^{-2b_i}$~\cite{orhan2015low} where $b_i$ is the number of quantization bits for each real and imaginary part of $\tilde y_i$.
The MSQE of $\tilde y_i$ with $b_i$ quantization bits is
\vspace{-0.5 em}
\begin{align}
\label{eq:msqe}
D_i(b_i) 
&= \mathbb{E}[|\tilde{y}_i - \tilde{y}_{{\rm q}i}|^2]= \frac{\pi\sqrt{3}}{2}\sigma_i^2\,2^{-2b_i}
\end{align}
In the next subsection, we solve the BA problem by minimizing the total MSQE subject to the power constraint.

\vspace{-0.5em}
\subsection{Power Constrained Bit Allocation}
\label{subsec:B}

We assume that the power consumption at each ADC $P_{\rm ADC}$ scales exponentially in the number of bits per sample $b$ as~\cite{orhan2015low} 
\begin{align}
\label{eq:ADCpower}
P_{\rm ADC}(b)= cW2^{b}
\end{align}
where $c$ is the energy consumption per conversion step (e.g. $494\ {\rm fJ}$) and $W$ is the sampling rate.
Noting that $b$ is a non-negative integer, the MSQE minimization can be formulated as an integer optimization problem.
To find a closed-form solution, we relax the integer problem with ${\bf b} \in \mathbb{Z}^N_+ $ to the real number problem with ${\bf b} \in \mathbb{R}^N $. 
Despite the fact that $P_{\rm ADC}(b) = 0 \text{ for } b \leq 0$, we also relax the problem by considering $P_{\rm ADC}(b) = cW2^b \text{ for } b \in \mathbb{R}$.
Having the power of $N$ $\bar b$-bit ADCs as the constraint, the minimization problem is
\vspace{-0.5 em}
\begin{align}
\label{eq:opt_power}
&\qquad \mathbf{\hat{b}} 
= \argmin_{\mathbf{b}=[b_1,\cdots,b_N]^\intercal} \sum_{i=1}^{N}D_i(b_i)\\ \nonumber
\text{s.t.} &\quad \sum_{i=1}^{N} P_{\rm ADC}(b_i) \leq NP_{\rm ADC}(\bar b),\ {\bf b}\in\mathbb{R}^{N}.
\vspace{-0.5em}
\end{align}
By defining $x_i = 2^{-2b_i}$, $\bar x = 2^{-2\bar b}$ and $v_i = \sigma_i^2$, we can convert~\eqref{eq:opt_power} into a simpler form given as
\begin{align}
\label{eq:trans_problem}
&\ {\bf \hat x}= \argmin_{{\bf x} = [x_1,\cdots,x_N]^\intercal} {\bf v^\intercal x} \\ \nonumber
\text{s.t.} \quad \sum_{i=1}^{N}&{x_i^{-\frac{1}{2}}}\leq N \bar x^{-\frac{1}{2}}  \text{, and } {\bf x} > {\bf 0}_N 
\end{align}
where ${\bf 0}_N$ is a $N \times 1$ zero vector. Note that~\eqref{eq:trans_problem} is the equivalent problem to~\eqref{eq:opt_power} and is a convex optimization problem. The global optimal solution of~\eqref{eq:opt_power} can be achieved by solving Karush\--Kuhn\--Tucker (KKT) conditions for~\eqref{eq:trans_problem}.
By relaxing ${\bf x >0}_N$ to ${\bf x \geq 0}_N$ with ${\bf h} = \begin{bmatrix}
\sum_{i=1}^{N}{x_i^{-\frac{1}{2}}}- N \bar x^{-\frac{1}{2}} \\ 
 -{\bf x}
\end{bmatrix}$, KKT conditions become
\vspace{-.5 em}
\begin{align} 
\label{eq:kkt1}
&{\bf v} + J({\bf x})^\intercal{\pmb \mu} = {\bf 0}_N \\ \label{eq:kkt2}
&\mu_i\,h_i = 0,\quad \forall\, i \in \{1,\cdots, N+1\}\\ \label{eq:kkt3}
&\mathbf{h} \leq \mathbf{0}_{(N+1)}\\ \label{eq:kkt4}
&{\pmb \mu} \geq \mathbf{0}_{(N+1)}
\end{align}
where the Jacobian matrix of $\mathbf{h}$ is 
$J(\mathbf{x}) = \begin{bmatrix}
       \mathbf{p}
 &      -\mathbf{I}_{N}
\end{bmatrix}^\intercal$, the vector ${\bf p} = \left[-\frac{1}{2}x_1^{-\frac{3}{2}}, \cdots,-\frac{1}{2}x_N^{-\frac{3}{2}} \right]^\intercal$, and ${\pmb \mu} \in \mathbb{R}^{(N+1)}$ is the vector of the Lagrangian multipliers. 
As $x_i \neq 0,\ \forall\, i \in \{1, \cdots, N\}$, the Lagrangian multipliers become $\mu_j = 0,\ \forall\, j \in \{2, \cdots, N+1\}$ from~\eqref{eq:kkt2}.
So, \eqref{eq:kkt1} guarantees $\mu_1 \neq 0$ as ${\bf v \neq 0}_N$, and~\eqref{eq:kkt2} gives $h_1 = 0$ meaning that the equality holds for the power constraint.
From~\eqref{eq:kkt1} and~\eqref{eq:kkt2}, we have
$v_i = \frac{1}{2}x_i^{-\frac{3}{2}}\mu_1, \text{ and }\sum_{i=1}^{N}{x_i^{-\frac{1}{2}}}= N \bar x^{-\frac{1}{2}}$,
which gives $\mu_1 =\left\{ \frac{\bar x^{\frac{1}{2}}}{N}\sum_{j=1}^{N}(2v_j)^{\frac{1}{3}}\right\}^3 > 0$.
Putting $\mu_1$ into $v_i = \frac{1}{2}x_i^{-\frac{3}{2}}\mu_1$, the solution becomes 
$x_i = \bar x \left\{ \frac{1}{N}\sum_{j = 1}^{N}{\left(\frac{v_j}{v_i}\right)}^{\frac{1}{3}}\right\}^2$.
From the definitions of $x_i, \bar x \text{ and }v_i$,
\begin{align}
\label{eq:opt_BA}
 \hat b_i = \bar b - \log_2\left(\frac{1}{N}\sum_{j = 1}^{N}\left\{\frac{1+{\rm SNR}^{\rm RF}_j}{1+{\rm SNR}^{\rm RF}_i}\right\}^{\frac{1}{3}}\right)
\end{align}
where ${\rm SNR}_j^{\rm RF} = \frac{\|[{\bf G}]_{j,:}\|^2}{N_0}$.
The allocated bits for the $i$th ADC increases logarithmically with $(1+\text{SNR}_i^{\rm RF})^{1/3}$ and decreases logarithmically with the sum of  $(1+\text{SNR}_j^{\rm RF})^{1/3}$, $j = 1,\cdots, N$.
The near optimal BA can be achieved by mapping~\eqref{eq:opt_BA} into non-negative integers.
\vspace{-0.7em}
\subsection{Bit Allocation Algorithm}
\label{sec:algorithm}
\vspace{-0.3em}
\begin{algorithm}[!b]
\vspace{-0.5em}
\caption* {{\bf Algorithm} \ Near Optimal BA Algorithm for $N$ Antennas}
\label{Power}
\begin{enumerate}
\item Set power constraint $P_{\rm max} = NP_{\rm ADC}( \bar{b})$ using~\eqref{eq:ADCpower}
\vspace{-0.7em}
\item Set $\mathbb S = \{1 \ldots N\}$ and $P_{\rm total} = 0$
\vspace{-0.7em}
\item {\bf for} $i = 1 \ldots N$
\vspace{-0.7em}
  \begin{enumerate}
  \item Compute $\hat{b}_i$ using~\eqref{eq:opt_BA} and $b_i = \max(0, \lceil \hat{b}_i \rceil)$
  \item {\bf if} ($b_i = 0$), $\mathbb S = \mathbb S - \{i\}$
  \item {\bf else} $p_i = P_{\rm ADC}( b_i )$ and $P_{\rm total} = P_{\rm total} + p_i$ 
\begin{enumerate}
\item [$\circ$]{\bf if} ($\hat b_i \in \mathbb N$), $\mathbb S = \mathbb S - \{i\}$
\end{enumerate}
  \end{enumerate}
\vspace{-0.7em}
\item {\bf if} $P_{\rm total} \le P_{\rm max}$, {\bf  return b}
\vspace{-0.7em}
\item {\bf for} $i \in \mathbb S$, $T_i = T_{\rm rel}(i)$ using~\eqref{eq:Trel}
\vspace{-0.7em}
\item {\bf while} $P_{\rm total} > P_{\rm max}$ \label{while}
\vspace{-0.7em}
  \begin{enumerate}
  \item $i^{*} = {\rm argmin}_{i \in \mathbb S} T_i$
  \item $b_{i^{*}} = {b}_{i^{*}} - 1$ and $\mathbb S = \mathbb S - \{i^{*}\}$
  \item $P_{\rm total} = P_{\rm total} - p_{i^*} + P_{\rm ADC}(b_{i^*})$
  \end{enumerate}
\vspace{-0.7em}
\item {\bf return b}
\end{enumerate}
\vspace{-0.7em}
\end{algorithm}

To perform the non-negative integer mapping, $\hat b_i \leq 0$ is mapped to $0$, i.e., the ADC pairs with $\hat b_i \leq 0$ are inactivated. 
 This mapping does not violate the actual power constraint as $P_{\rm ADC}(b) = 0 \text{ for } b \leq 0$.
 Next, non-integer $\hat b_i >0$ is mapped to $\lceil \hat b_i \rceil$. 
 If $\sum_{i\in \{i | \hat b_i >0 \}}P_{\rm ADC}(\lceil \hat b_i \rceil) > NP_{\rm ADC}(\bar b)$, we need to map some non-integer $\hat b_i$ to $\lfloor \hat b_i \rfloor$, which reduces the power consumption while increasing the MSQE.
The $\lfloor \hat b_i \rfloor$ mapping can always satisfy the power constraint: (i) for $\hat b_i <0$, there is no power increase when mapping $\hat b_i < 0$ to $0$, and (ii) for $\hat b_i >0$, the total power becomes $\sum_{i\in \{i | \hat b_i >0 \}}P_{\rm ADC}(\lfloor \hat b_i \rfloor) \leq \sum_{i\in \{i | \hat b_i >0 \}}P_{\rm ADC}(\hat b_i)$.

Due to the non-linearity of the MSQE and power with respect to $\hat b$, we propose a trade-off function which represents the MSQE increase per unit power savings after mapping $\hat b_i$ to $\lfloor \hat b_i \rfloor$ for non-integer $\hat b_i >0$ as
\vspace{-0.5em}
 \begin{align}
 \label{eq:tradeoff}
 T(i) = \left|\frac{\Delta D_i(\hat b_i)}{\Delta P_{\rm ADC}(\hat b_i)}\right| =\left|\frac{D_i(\hat b_i) -  D_i(\lfloor \hat b_i \rfloor) }{P_{\rm ADC}(\hat b_i)-P_{\rm ADC}(\lfloor \hat b_i \rfloor)}\right|.
\vspace{-0.5em}
 \end{align}
Then, $\hat b_i$ with the smallest $T(i)$ is re-mapped to $\lfloor \hat b_i \rfloor$ for a better trade-off of quantization error vs. power consumption.
This repeats for $\hat b_i$ with the next smallest $T(i)$ until the power constraint is satisfied.
For a practical BA algorithm, we use a relative trade-off function obtained from~\eqref{eq:tradeoff} as
\begin{align}
\label{eq:Trel}
T_{\rm rel}(i) = \frac{2^{-2\lfloor\hat b_i\rfloor}-2^{-2\hat b_i}}{2^{\hat b_i}-2^{\lfloor \hat b_i\rfloor}}\sigma^2_i.
\end{align}
Algorithm shows the described BA algorithm. 
Since the solution~\eqref{eq:opt_BA} is in closed-form and while loop at line~\ref{while} occurs a maximum $N$ times, the computational complexity is $\mathcal O(N^2)$, and while loop at line~\ref{while} will always end as explained.

\section{Simulation results}
\label{sec:simulation_results}

To evaluate the proposed BA method, we generate the mmWave channel based on~\cite{thomas20143d}. 
For a narrowband mmWave channel, we assume 1 cluster and 4 subpaths $(p = 4)$ between each user and the BS, and the system uses a $73$ GHz carrier frequency with the antenna spacing of $\lambda/4$.
The average EVM is adopted as a performance measure; $\text{EVM} \ (\%)= \frac{\|\bx-\hat\bx\|}{\|\bx\|}\times 100$, where $\bx$ is the vector of the quadrature phase shift keying symbols 
from $M$ users and $\hat\bx$ is the decoded signal vector using a zero-forcing equalizer. 
Defining SNR$=\frac{\mathbb{E}[|x_i|^2]}{N_0}$,
we provide link-level simulation results of the near optimal BA comparing with the non-quantization and low-resolution quantization cases. 

\begin{figure}[!t]
\centering
$\begin{array}{c}
\mbox{(a)}\\
{\resizebox{0.86\columnwidth}{!}
{\includegraphics{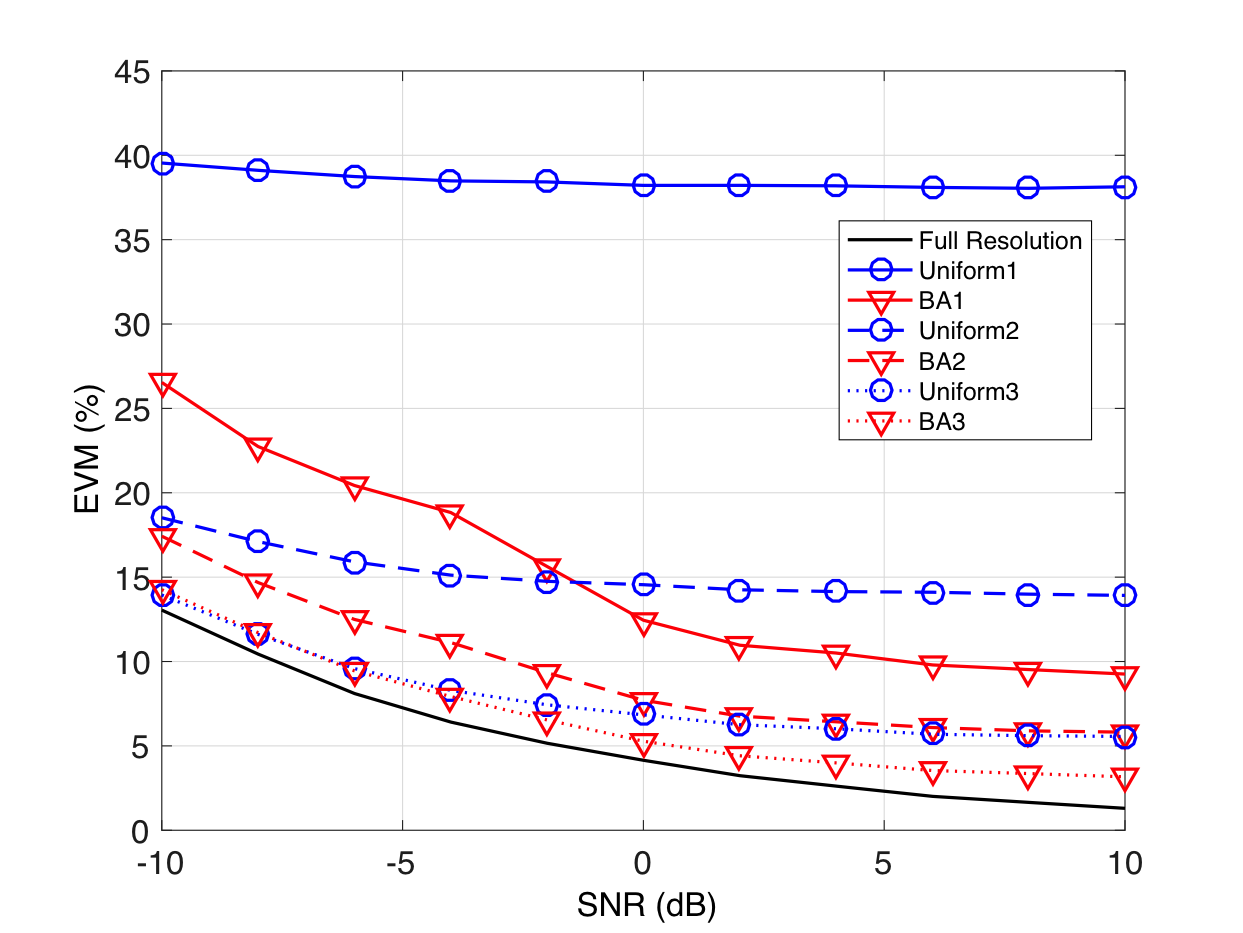}}}  \\
\mbox{(b)}\\
{\resizebox{0.86\columnwidth}{!}
{\includegraphics{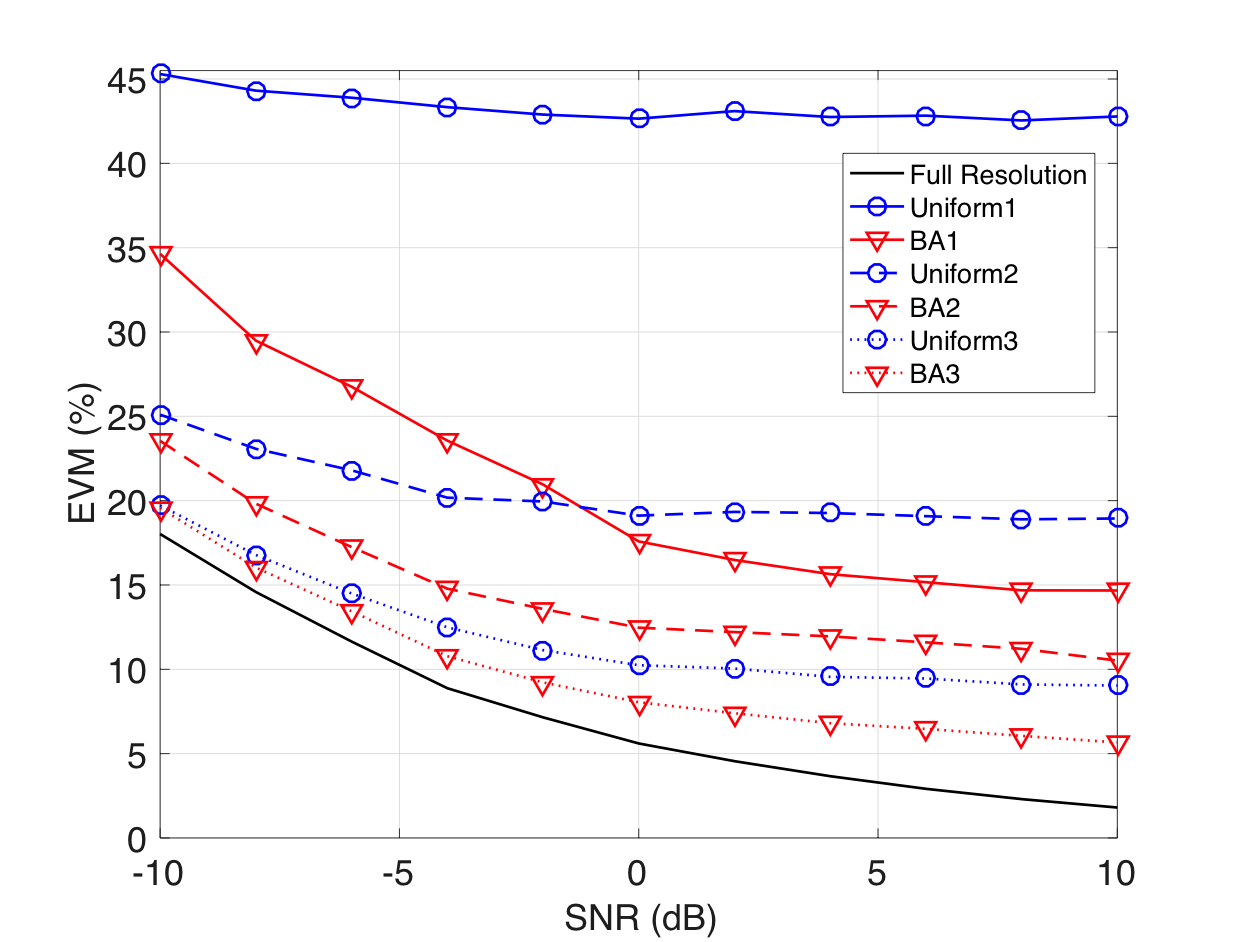}}} 
\end{array}$
\caption{
The average EVM of the near optimal BA algorithm and corresponding uniform bit ADCs, $\bar b \in \{ 1, 2,3\}$ for (a) $8$ users and (b) $16$ users with $256$ BS antennas.
} 
\label{fig:2}
\vspace{-1 em}
\end{figure}

\fig{fig:2} shows the average EVM for $N = 256$ BS antennas with (a) $M = 8$ and (b) $16$ users, respectively.
Full resolution indicates a non-quantization case considered as an ideal case for benchmark.
Uniform1, Uniform2 and Uniform3 are the cases of 256 antennas equipped with $\bar b$-bit ADCs with $\bar b = 1, 2 \text{ and } 3$ respectively.
BA1, BA2 and BA3 are the results of the proposed BA algorithm that correspond to Uniform1, Uniform2 and Uniform3; i.e., BA1, BA2 and BA3 satisfy the power constraint with $\bar b = 1, 2 \text{ and } 3$. 

 Unlike the uniform bit ADC cases, some BA cases have inactive (0-bit) ADCs while assigning more bits to other ADCs with larger aggregated channel gains.  
For example, BA1 with 8 and 16 users turns off 134 and 167 ADCs, respectively, out of 256 ADCs when the SNR is 10 dB.
Accordingly, the proposed BA method outperforms the low-resolution quantization case for both $8$ and $16$ users when using 256 antennas.
In particular, our BA algorithm shows a larger EVM gap with a smaller $\bar b$.
This illustrates that our algorithm is more effective for the communication environment with a harsher power constraint.
The EVM gap also increases with increasing SNR as smaller noise allows the proposed algorithm to more effectively reduce the quantization error of desired signals.
Beyond SNR $\approx -1$ dB,  BA1 shows lower EVM even than Uniform2 which consumes twice the power of BA1 for both  $8$ and $16$ users.
Beyond SNR $\approx 2$ dB, BA2 shows similar EVM as Uniform3 which consumes twice the power of BA2 for $8$ users.
Thus, our BA algorithm can achieve a large improvement in the communication performance compared to a conventional low-resolution quantization strategy.

%
%

\section{Conclusion}
\label{sec:conclusion}

In this paper, the primary contribution is a near optimal low-complexity bit allocation technique for mmWave MU-massive MIMO uplink systems. 
Adopting RF preprocessing, the proposed algorithm projects the received signals onto the beamspace by spatial Fourier transform to exploit mmWave channel sparsity. 
Assuming an additive quantization noise model, we formulate the relaxed MSQE minimization problem under a power constraint.
Through a non-negative integer mapping, the solution of the problem is used to allocate bits at each ADC.
In simulation, the proposed method achieved large EVM improvement when compared to the low-resolution quantization with equal or less power and performed more effectively under a harsher power constraint.





\clearpage

\bibliographystyle{IEEEtran}
\bibliography{ICASSP_HybridBA.bib}

\end{document}